\documentclass[aps,preprint,showpacs,showkeys]{revtex4}
\usepackage{graphicx}
\usepackage{amssymb}
\usepackage{bm}
\begin{document}
\setcounter{page}{1}
\title[]{Fabrication of Analog Electronics for Serial Readout of Silicon
Strip Sensors}
\author{E. \surname{Won}}
\email{eunilwon@korea.ac.kr}
\thanks{Fax: +82-2-927-3292}
\author{J. H. \surname{Choi}}
\author{H. \surname{Ha}}
\affiliation{Department of Physics, Korea University, Seoul 136-713, Korea}
\author{H. J. \surname{Hyun}}
\author{H. J. \surname{Kim}}
\author{H. \surname{Park}}
\affiliation{Department of Physics, Kyungpook National
University, Daegu 702-701, Korea}
\date[]{Received January 5 2006}

\begin{abstract}
A set of analog electronics boards for serial readout of silicon 
strip sensors was fabricated. 
A commercially available amplifier is mounted on a
homemade hybrid board in order to receive analog 
signals from 
silicon
strip sensors. Also, another homemade circuit board is fabricated
in order to translate 
amplifier control signals into a suitable format and to provide bias voltage to
the amplifier as well as to the silicon sensors. We discuss technical
details of the fabrication process and performance of the circuit
boards we developed. 
\end{abstract}

\pacs{83.85.Gk, 84.30.Le, 84.30.Sk}

\keywords{amplifier,electronics, ASIC}

\maketitle

\section{INTRODUCTION}
 Over the last thirty years, there have been impressive developments
in silicon strip sensors and their readout electronics
in the field of elementary particle physics.  
They were first used more than twenty years ago for heavy flavour searches
in fixed target experiments~\cite{fixed}. In particular, their potential 
use as
high precision vertex detectors around high energy colliders, both
for electron-positron and proton-antiproton 
machines, has initiated further
development of high performance semiconductor detectors 
till now~\cite{collider_delphi,collider_belle,collider_cdf}. 
Subsequently, it became clear that much higher density electronics
was required and it drove the construction of integrated circuit
amplifiers in metal-oxide-silicon technology. Therefore,  
application specific integrated circuit (ASIC) technology has been 
heavily used in designing readout electronics
for silicon sensors in the particle physics experiments~\cite{hall,raible}. 
The interface electronics board between silicon sensors and readout 
ASIC chips is traditionally called hybrid boards and the 
experiment-specific hybrid boards have been produced for various 
experiments~\cite{raible,belle_hybrid,atlas}.  
The design of such hybrid boards should consider 
cooling system for collider experiments and low electrical noise performance for
detecting small signals.

 Recently, research activities on the high density readout electronics
have been extended to the field of  
high resolution
medical imaging~\cite{medical} 
as well as charged particle trackers in the future particle physics 
program~\cite{lctracker}.
Therefore, it becomes clear that the knowledge and 
experience in fabricating hybrid board and reading out analog signals
from it may be one of important items for the participation to such programs. 
In this respect,
we discuss development of several different types of
hybrid boards and related electronics board
with the technology available domestically.
Section~\ref{hybrid1}
describes our first prototype hybrid board. The fabrication of 
detector bias and dc voltage delivery for the operation of ASIC chip,
and the control logic translator system is described in section~\ref{plc}. 
We also fabricated a specialized hybrid board in order to test the
ASIC amplifier itself and it is described in section~\ref{hybrid2}.
Our latest design that mounts a 17 channel single-side silicon
detector is described in section~\ref{hybrid3}.

\section{DESIGN OF HYBRID BOARD I}\label{hybrid1}

\begin{figure}[t!]
\includegraphics[width=14cm]{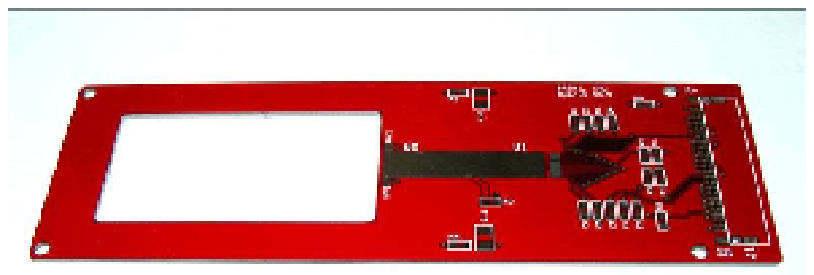}
\caption{
A picture of the hybrid board version 0.9. A large rectangular 
hole on left is prepared
for future beam or radioactive source tests when a sensor is mounted on
the hybrid board. U1 is the pad for the VA chip and U2 is for the $R/C$ chip.
} 
\label{fig:hybrid9.0}
\end{figure}

 In this section, we discuss the development of our first prototype of the
hybrid board that mounts a commercially available high density
ASIC amplifier, the VA chip~\cite{vachip1,vachip2,ideas}. 
It has in total 128 analog
channels and each channel contains a charge-sensitive
preamplifier, a shaper, a track-and-hold, and multiplexing capacity. 
In order to communicate to the VA chip, one has to wire-bond approximately 30 
lines 
of various analog and digital signals from the VA chip to
the hybrid board. Since the width of the VA chip
is 5 mm, the layout size of 30 pads on the 
printed circuit board (PCB) also should be in the
similar size in order to make good electric connections to the VA chip.
It turns out that a pad width and 
the pitch of pads both should be on the order of
100 $\mu$m on the PCB in order to make a good ultrasonic wire-bonding to the VA chip.
However,
most of domestic, small-size vendors expressed 
difficulty in fabricating such fine structure PCB layout with their facility. 
Figure~\ref{fig:hybrid9.0} shows our first attempt to fabricate high 
density pads on the PCB from a domestic vendor. 
A large rectangular hole is made on the left corner of the board 
where the sensor is
to be mounted. This hole is placed in order to minimize the
material for the future radioactive source or beam tests. 
The PCB is made with four layers where analog and digital grounds are 
routed in the same layer.
One can also see a smaller, horizontally long rectangular pad (labeled as U2) near 
to the place for the VA chip (labeled as U1). This rectangular pad is for the $R/C$ 
chip~\cite{rcchip} 
as the detector
to be mounted at the design stage is a dc-type sensor~\cite{silicon}. 
This 
complicates the detector biasing method quite significantly because the $R/C$ chip we use
is known to break down at 70 V and therefore a voltage division is made to provide
full depletion bias voltage to sensors.

\begin{figure}[t!]
  \mbox{
   \includegraphics[width=6cm]{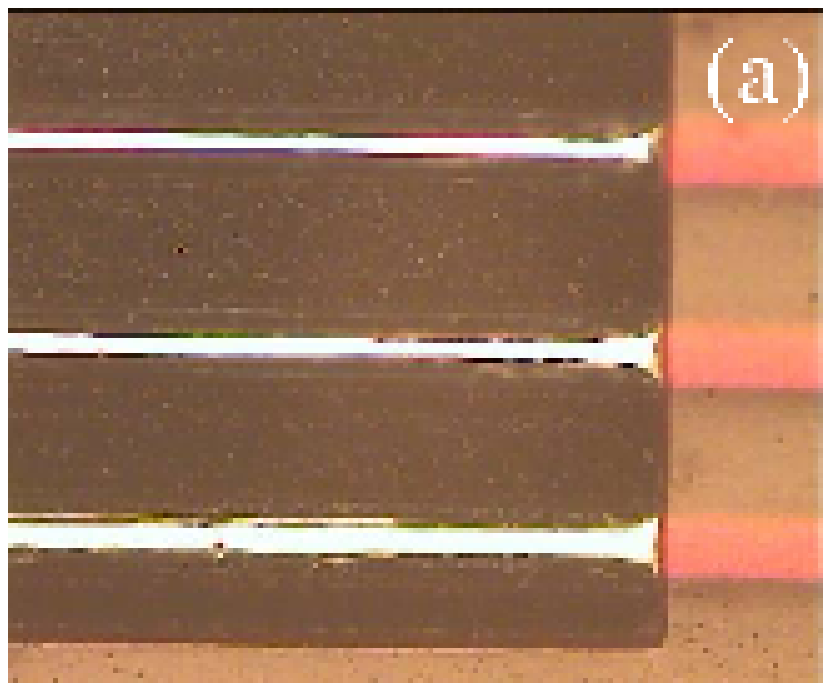} \quad
   \includegraphics[width=6cm]{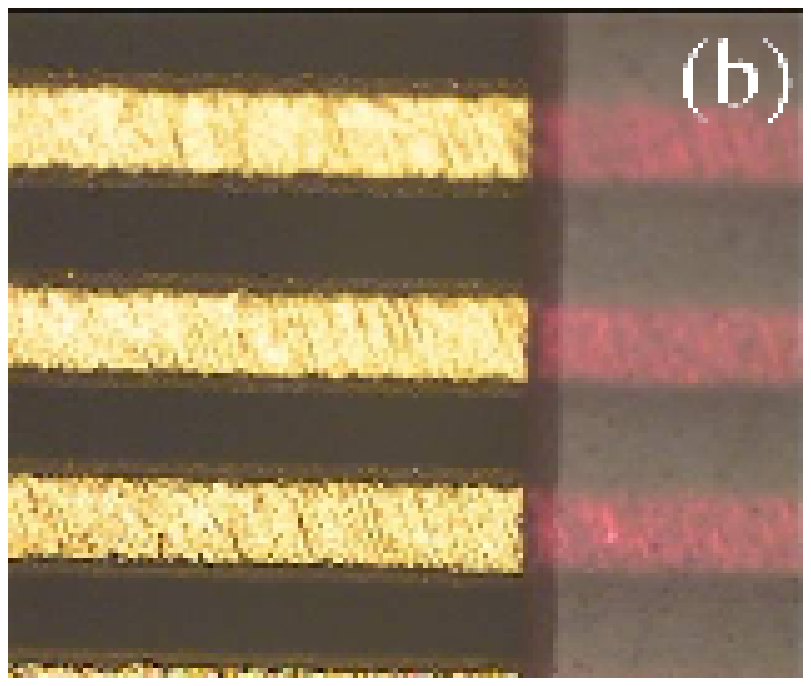} 
  }
  \mbox{
   \includegraphics[width=6cm]{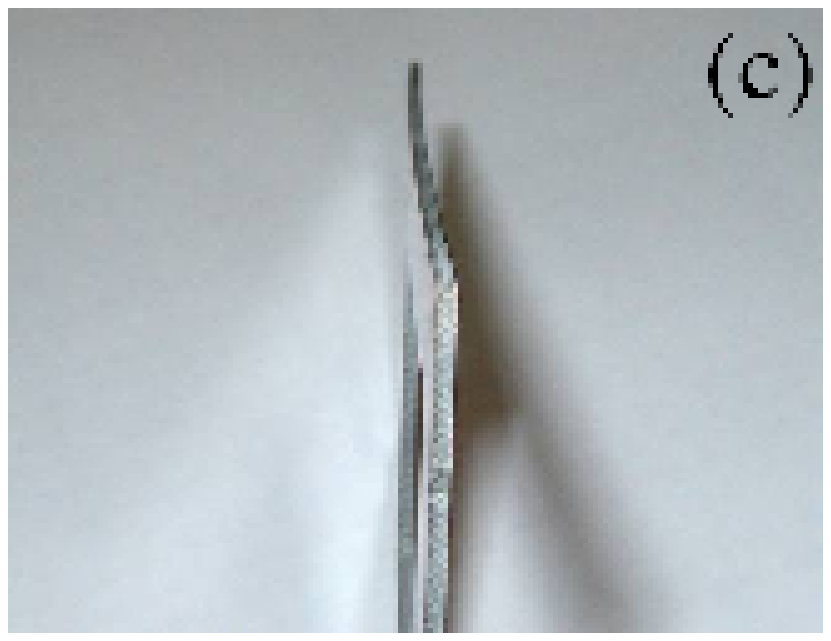} \quad
   \includegraphics[width=6cm]{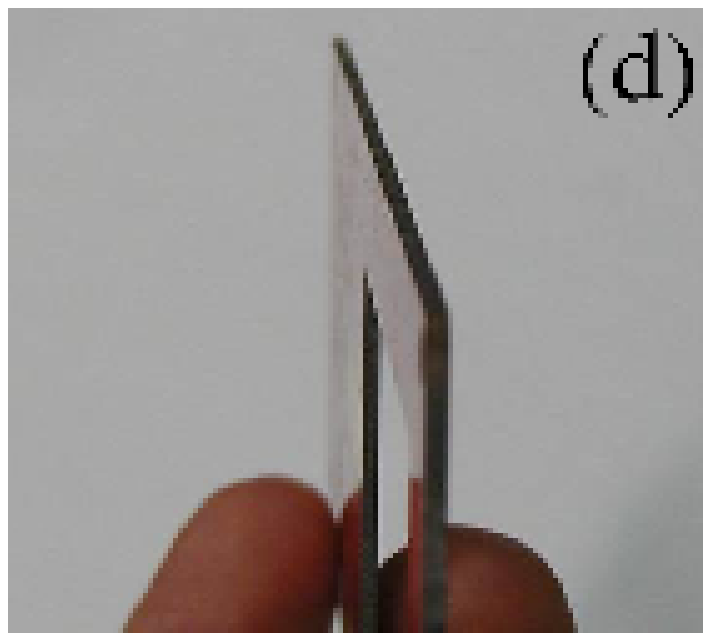} 
  }
\caption{
Zoomed views of hybrid board version 0.9 and version 1.0. Microscope picture of the
bonding pad on the board is shown in (a) for the version 0.9 and in (b) for 
the version 1.0.
They are in same scale and clear improvement in the quality of the
layout can be seen in (b). Side views of the boards are shown in (c) for
the version 0.9 and in (d) for the version 1.0. 
} 
\label{fig:pitch}
\end{figure}

A tin-lead alloy was used in order to
cover all copper pads on the 
PCB. Later we realized that in some countries there are at least 
directives restricting the use of lead for such purpose and therefore we 
abandoned
the use of tin-lead alloy completely.
The use of tin-lead alloy resulted in significantly
bad quality in the
layout of the pad outlines. A microscope picture of the pads in
Fig.~\ref{fig:pitch} (a) illustrates the situation.  
The three horizontal lines in the figure represent bonding pads on
the hybrid PCB. The average width of pads (thickness in vertical direction)
is always less than 15 $\mu$m 
and is too narrow for any practical
use. We labeled this first PCB board as the version 0.9. There is
another problem in the version 0.9. The board was not flat
and it prevented us from mounting 
silicon sensors as it does not provide mechanically stable
configuration.  This situation is also
shown in Fig.~\ref{fig:pitch} (c). After the fabrication of
the version 0.9, there has been series of discussion with
technicians from the vendor in order to identify source 
of these two problems. The problem with the poor quality of
the pad outline is partially solved by 
modifying one of chemical etching processes in their PCB fabrication. 
We also use gold to cover all copper pads on the PCB and it
partially helped in improving the quality of the bonding pad.
The source of the non-flat structure of the board was due to
improper handing of the PCB during the cooling process.
After identifying sources of troubles mentioned above, 
we fabricated our second prototype hybrid board with minor 
modification as far as the design is concerned. A placeholder
for a lemo connection to the analog signal output is made for debugging purpose
in the second prototype. Figure~\ref{fig:pitch}
(b) shows the quality of the pitch for the second prototype board.
Measurements showed that the width of pads is 110 $\mu$m which satisfies
our specification. The trouble with the non-flatness of the board is
also disappeared in the second prototype and it is clearly shown
in Fig.~\ref{fig:pitch} (d). We label the second prototype as the 
hybrid version 1.0.  
We note that in order this to be used in the real collider
environment, it has to deal with the heat generated during the
collision. One of solutions is to make the PCB with ceramic material
but we did not investigate the possibility of making ceramic PCB
at this time for a quick development of the board. 

\begin{figure}[t!]
  \mbox{
   \includegraphics[width=6cm]{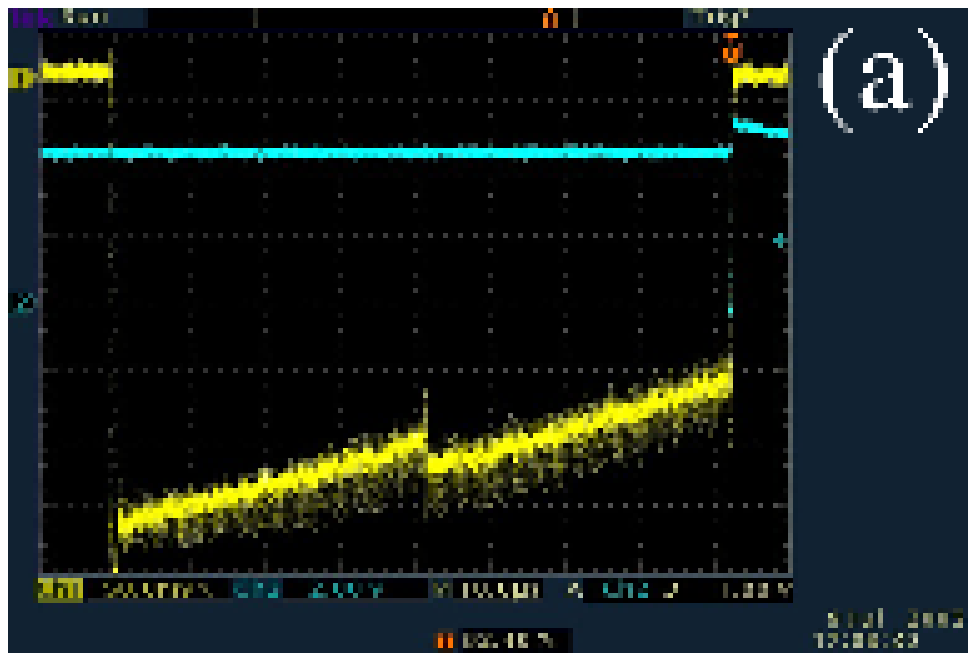} \quad
   \includegraphics[width=6cm]{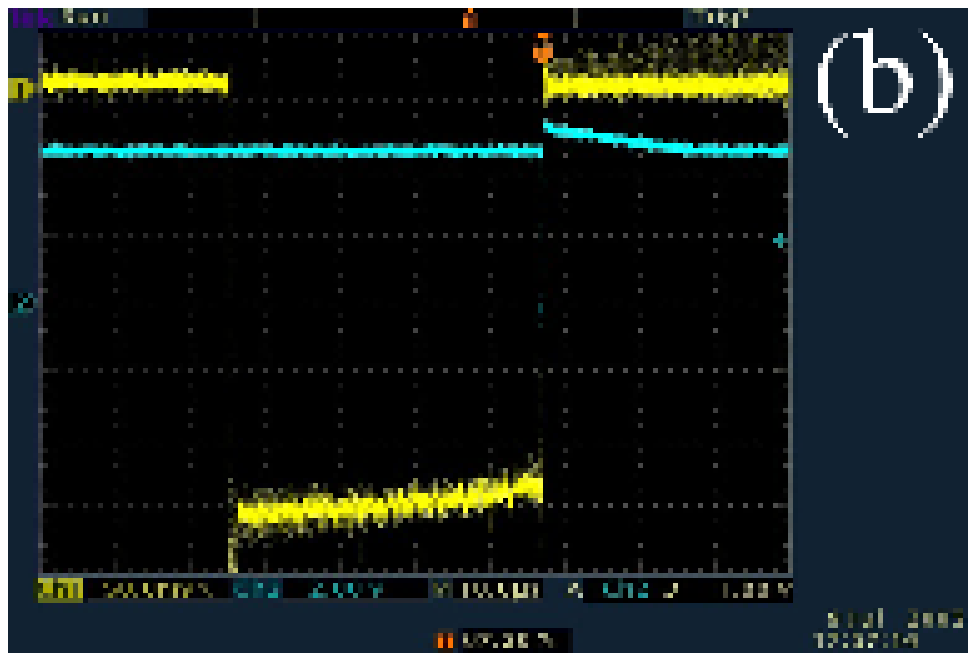} 
  }
  \mbox{
   \includegraphics[width=6cm]{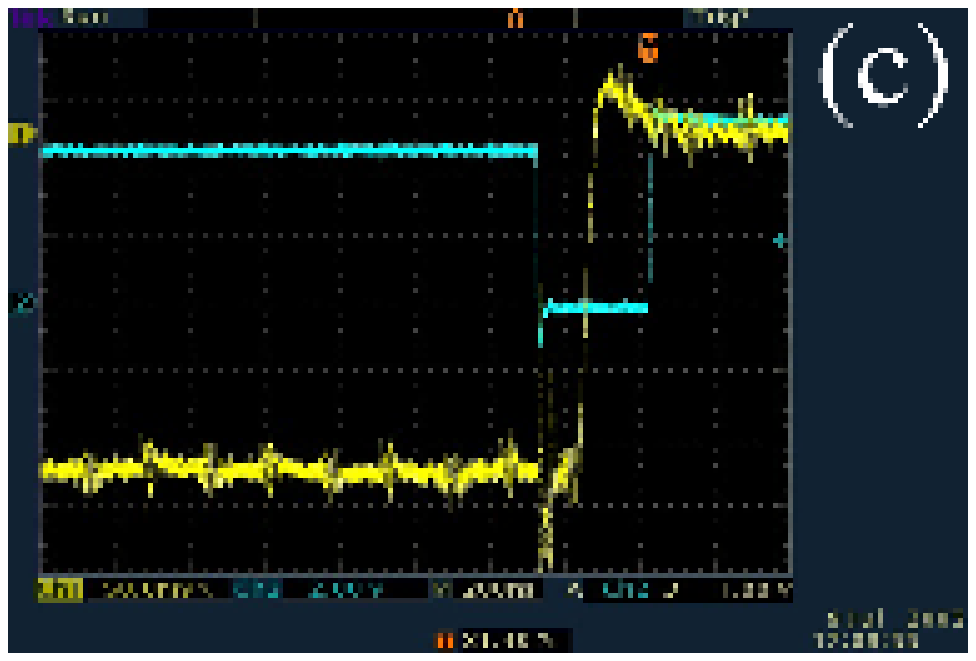} \quad
   \includegraphics[width=6cm]{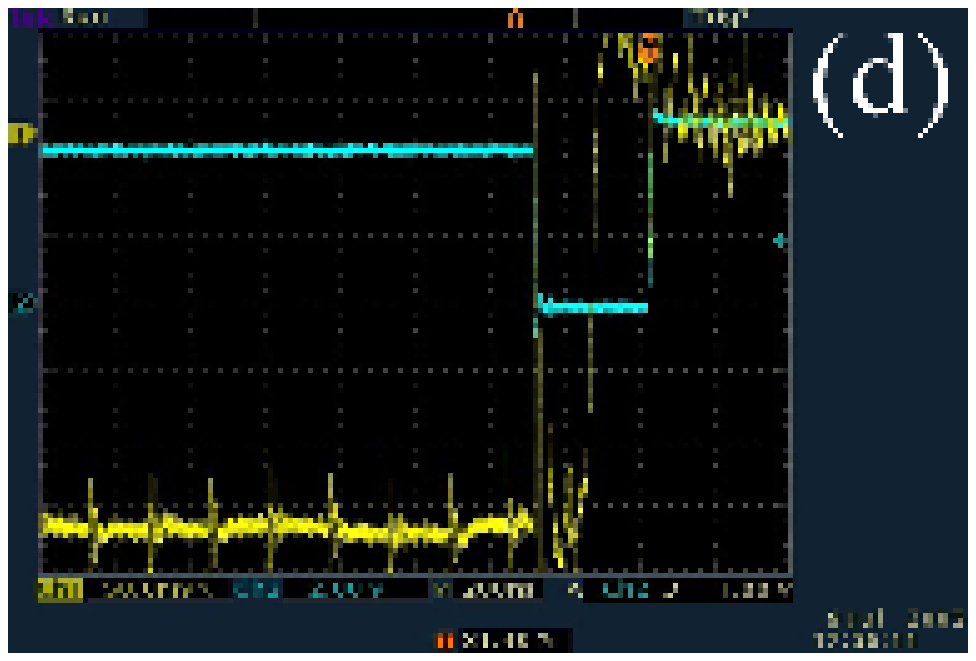} 
  }
\caption{
Analog and digital output signals from the VA chip captured in 
the oscilloscope. A distorted well-shape
signal in (a) and (b) are the analog outputs from the evaluation
board provided by the company and from our hybrid board version 1.0,
respectively. 
The evaluation board houses two VA chips and the width of the well
in (a) reflects it.
The other line in each figure is the digital control signal
({\sf shift\_out}) indicating the end of the serial readout. (c) and
(d) are zoomed views of (a) and (b) at the rising edge of the analog out.
} 
\label{fig:shiftout}
\end{figure}

 There are in total twenty passive surface-mounted components soldered
on the hybrid board. All components are 
a $F$-class which has $\pm$1 \% tolerance
from their specification values. 
A conductive epoxy from Chemtronics CW2400~\cite{chemtronics} 
is tested for a good ohmic contact with the gold pad on the
hybrid board and is used in order to mount
the VA chip on the hybrid board, as the bottom plate of the VA chip
requires an electric contact.
After through electrical tests, assembled hybrid boards are shipped
to a local company~\cite{lpelectronics} 
for an ultrasonic wire-bonding between the VA chip
and the hybrid board. After the wire-bonding, the hybrid boards are
delivered back to the laboratory and a readout setup is made to
communicate with the hybrid board. A dc power supply is connected
to a homemade electronics board in order to provide voltage and
current sources to the hybrid board. We discuss the detailed design
of this second homemade 
board in the following section. The control signals are
generated from a commercially available 
field programmable gate array (FPGA) test board from Xilinx~\cite{xilinx}.
It has a SPARTAN XC3S200 on the board and a 
very high speed integrated circuit hardware description language 
(VHDL) firmware 
is written by us
in order to generate LVCMOS control logic signals to be sent to the VA chip.
The detailed time structure of these control signals may be
found from the reference~\cite{vachip2}.
The indication of a successful communication with the VA chip may
be the presence of a return signal from the VA chip. To be more specific, 
when all 128
channels are serially
read out, there is a signal coming from the VA chip, indicating
a serial data readout is completed. In the reference~\cite{vachip2}, it 
is referred as
{\sf shift\_out} and we confirmed that we were able to see this 
line became active-low, immediately after all  128 channels were read out. 
Figure~\ref{fig:shiftout} (a) and (b) show the behavior of 
the analog output and the {\sf shift\_out} signals, captured
in an oscilloscope from a commercially available VA evaluation
board~\cite{ideas} that was tested by us, 
and from our homemade hybrid board version 1.0. The well-like
signals in Fig.~\ref{fig:shiftout} (a) and (b) show the analog outputs from the
evaluation board and our hybrid board version 1.0, respectively.
Since the evaluation board we purchased 
has two VA chips on the board, the
width of the well from the evaluation board in Fig.~\ref{fig:shiftout} (a)
is twice larger than the width from the hybrid board version
1.0 in Fig.~\ref{fig:shiftout} (b). At this stage, both boards have no sensors
mounted and therefore the analog outputs are pedestals only.
The other signals in Fig.~\ref{fig:shiftout} (a) and (b) are {\sf shift\_out}
and should be active-low at the end of the serial readout 
of the VA chip. Such behavior can be clearly seen
from the zoomed view in Fig.~\ref{fig:shiftout} (c) for the evaluation
board and (d) for the hybrid board version 1.0, respectively. 
It appears that 
cross-talk from the clock signal at the edge 
to the analog output signal is somewhat
worse in the hybrid board 1.0, 
as indicated in 
Fig.~\ref{fig:shiftout} (c) and (d). 
We attribute that it is originated from
the ground
routing issues in the PCB design or imperfect impedance matching but
no conclusive statement can made at this moment without further study.

\section{DESIGN OF POWER, LOGIC TRANSLATORS and CURRENT SOURCES}\label{plc}
\begin{figure}[t!]
\includegraphics[width=14cm]{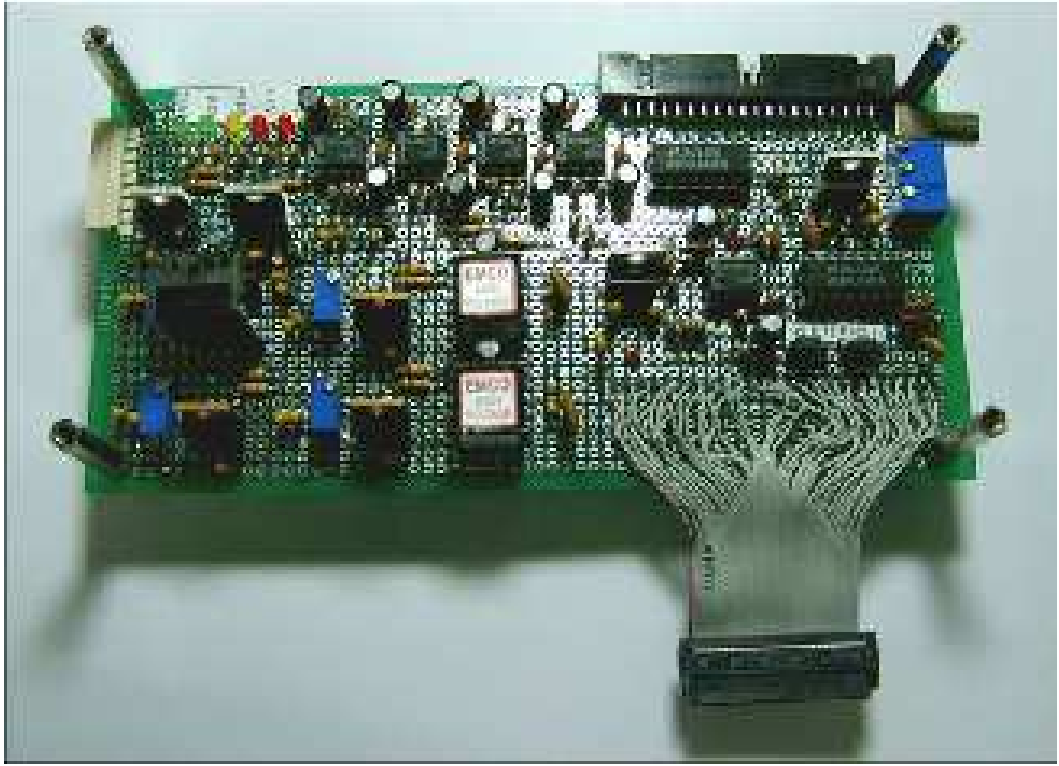}
\caption{
A picture of the PLC board prototype with hand-wirings. A micro-connector 
at the bottom right is for the hybrid board, the one on the top-right is
for the commercial evaluation board with a FPGA, and the one on the
top left is for dc power connection.
} 
\label{fig:plc_hand}
\end{figure}

In order to operate the VA chip, one has to provide several
voltage and current sources, and non-standard control logic
signals for serial readout and calibration purposes.
Also, bias voltage for the silicon sensors has to be 
provided as well. In order to provide 
power, logic translators, and current sources (PLC), we
developed another homemade electronics called PLC board.
We started with a hand-soldered prototype which is
shown in Fig.~\ref{fig:plc_hand}. There are four dc power lines connected
to the PLC board: $\pm$6.6 V for the main power that operates components
mounted, $+$4 V for the sensor bias voltage, and $+$16 V for the extra
bias for p-stop in the sensor we were planning to mount at the design stage.
In order to bias the silicon sensors, a dc to dc converter is designed using 
the EMCO high voltage chip Q01-05~\cite{emco}.
This model was chosen in order to provide positive and
negative voltages simultaneously
due to the fact that the $R/C$ chip was used in the
hybrid side.  The VA chip control signals are originally generated
from the outside of the PLC board and are fed into the PLC board as LVCMOS
logic. 
The PLC receives the control signals and translate them into a new logic with 
logical one begin $+$1.5 V and zero $-$2.0 V. 
Once it is done, signals 
are transferred to the hybrid board. Another functionality in the PLC board
is that the differential analog output from the VA chip is changed to a
single-ended signal through an analog receiver. This may be the source of
the noise that appears in following sections.
 
\begin{figure}[t!]
\includegraphics[width=14cm]{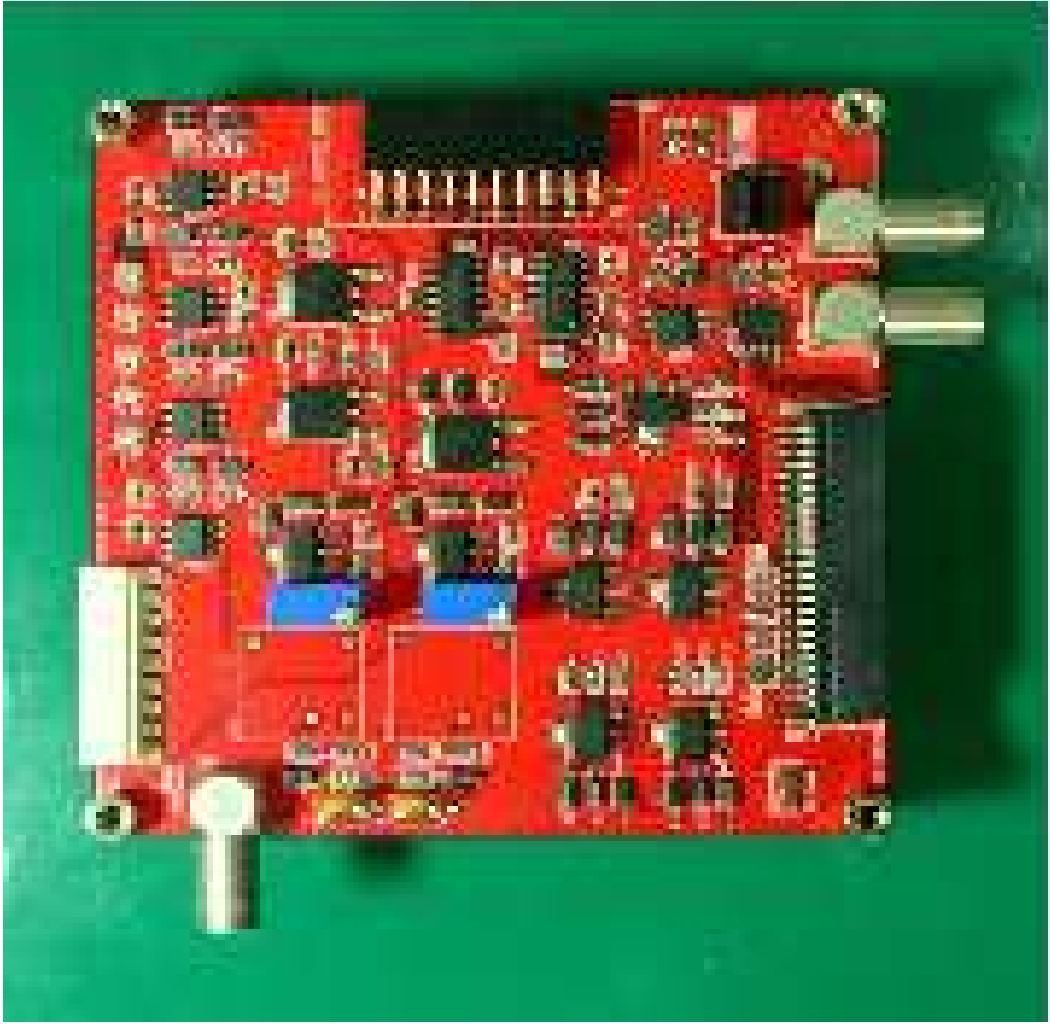}
\caption{
A picture of the PLC board fabricated with a
standard PCB process. The  physical size is 84 $\times$ 84 mm. 
A micro-connector 
at the bottom right is for the hybrid board, the one on the top is
for the commercial evaluation board with a FPGA, and the one on the
left is for dc power connection. 
All the components are chosen to be surface mountable in order to
reduce the size of the board.
} 
\label{fig:plc}
\end{figure}

 After careful studies on this prototype, a PCB is fabricated and
a picture of it
is shown in Fig.~\ref{fig:plc}. It has 6 layers and most of the
components are chosen to be surface mountable type, in order to reduce
the size of the board. The physical dimension is 84$\times$84 mm.
The analog and digital powers are now separated in the
PCB version of the PLC board and it enables us to reduce the noise due to the
digital clock. There are two square layouts on the PCB which are left blank
in Fig.~\ref{fig:plc}. Two EMCO high voltage chips are mounted on the back
side of the board due to the mistakes in the design of the PCB.

\begin{figure}[t!]
\includegraphics[width=14cm]{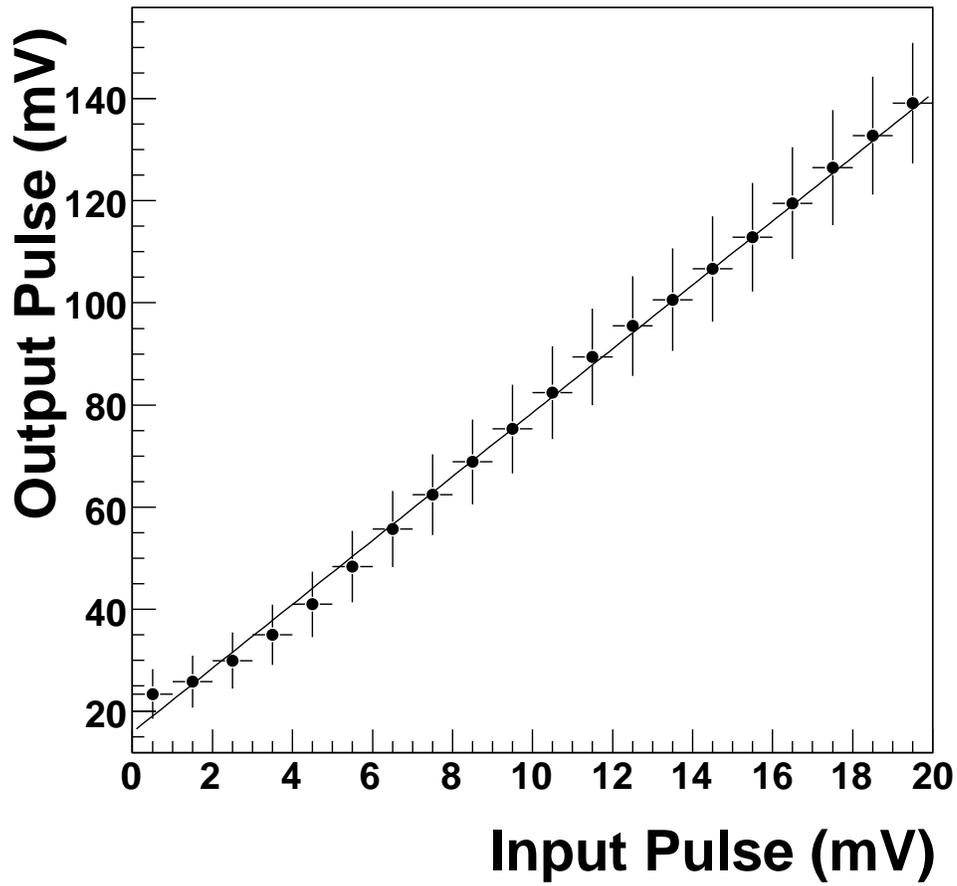}
\caption{
A response of the VA chip and hybrid board version 1.0 on test pulses
when the VA chip is in the calibration mode. Here, an input test pulse of 
3 mV corresponds to 1 MIP signal. 
} 
\label{fig:linearity}
\end{figure}

 With this PLC board, the VA chip is tested in a calibration mode. An
external coupling
capacitor is connected to the calibration input and test pulses are generated
in order to store electric charge to the capacitor. 
The channel to be tested is selected prior to the charge injection and
the amplified signal comes out without serialization of the data. In 
this sense, the calibration is somewhat different from the 
serial signal readout
from the sensor. 
Figure~\ref{fig:linearity}
shows the response of the VA chip for different test pulse values in mV.
In this test, one MIP corresponds to 3 mV. A good linearity is achieved
up to 7 MIPs, indicating a good performance of the VA chip with our assembled electronics. 
According to the VA chip specification, the dynamic range
reaches to $\pm$ 10 MIPs but the goal of our study is to develop the 
electronic boards and therefore we did not test the full dynamic range of the
VA chip in this
study.

\section{DESIGN OF HYBRID BOARD II}\label{hybrid2}

\begin{figure}[t!]
\includegraphics[width=14cm]{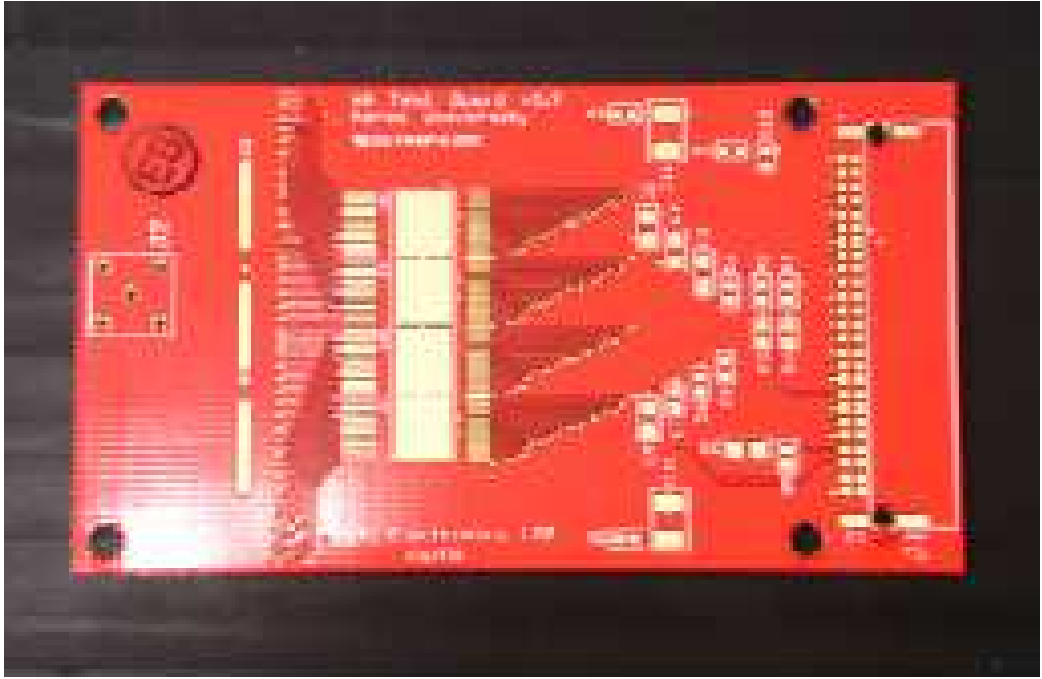}
\caption{
A picture of the VA-test board. This board is designed to mount in total
four VA chips. Selected channels 
in the VA side can be wire-bonded to pads on the PCB for a direct delivery
of current signals from an external pulse generator.
} 
\label{fig:vtest}
\end{figure}

\begin{figure}[t!]
\includegraphics[width=14cm]{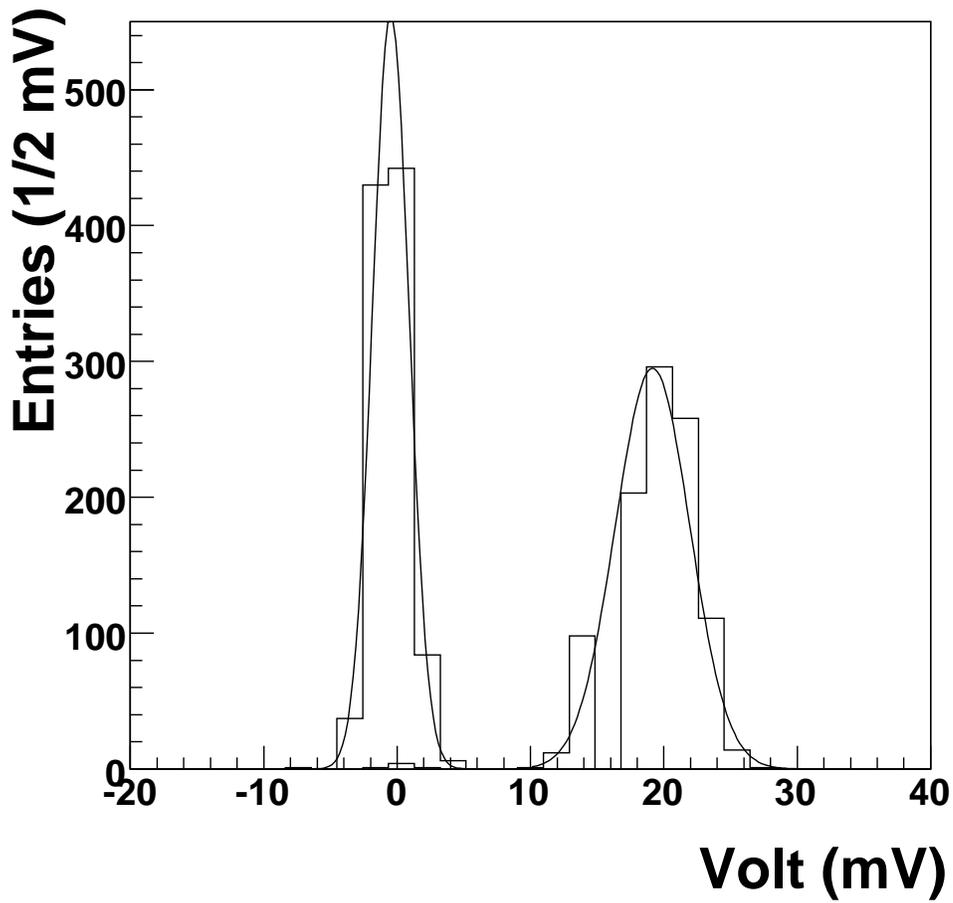}
\caption{
%% sn figure
Distribution of the measured analog output from the VA-test board when
a test pulse of one MIP signal is injected. The signal distribution
is on the right and the pedestal peak is shown on the left.
The measurement was
done directly from the oscilloscope by measuring the voltage outputs
from the hybrid board.  
} 
\label{fig:sn}
\end{figure}

 Due to the fact that the delivery of sensors are behind the schedule, we
decided to design another hybrid board that allows us to test the readability of 
the VA chip without real sensors. We label this one as VA-test hybrid
board. In this board, the rectangular hole for the sensor mounting is removed,
as indicated in Fig.~\ref{fig:vtest}. Instead, we place a set of wire-bond
pads on the board near to the input sides of the VA chips to be mounted.
One may see such configuration in Fig.~\ref{fig:vtest}, left side from the
layouts for the VA chips.
And then, direct wire-bonding from these pads to the input pads on the VA
chip is made in order to inject electric charges to the VA chip. This
is practically same method as the charge injection using 
the real sensor attached to the VA chip. This method is however somewhat 
different from the calibration mode mentioned in the previous section
because in the calibration process, there is no holding of the 
charge inside of the VA chip and no serialization of the data is carried out.
A lemo connector is prepared in order to inject electric charges 
using an external
pulse generator and channel selection is made through hand-soldering
to the pads to be tested, one at the time. Using this technique, one MIP
``signal'' is generated and the measured voltage output 
is shown in Fig.~\ref{fig:sn}. The
bump on the left corresponds to the pedestal of the entire electronics and 
the other bump on the right is one MIP signal. The measurement was
done directly from the oscilloscope by measuring the voltage outputs
from the hybrid board.  From the fits to two
bumps, the signal-to-noise ratio was measured to be 14. This is worse than
the nominal values one may get with the silicon sensors. One of the reasons
may be due to the fact that the electronics noise is larger. We discuss
it in detail in the next section. The equivalent noise charge (ENC) is measured to be
1740 $e^-$ ENC with
an 1 pF coupling capacitor and again this is a significantly worse value
than the value in the specification, 180 $+$ 7.5/pF $e^-$ ENC~\cite{ideas}. 
We discuss a possible reason for this in the next section.

 In this VA-test board, we also tested serial readout of multiple VA chips.
For this test, two VA chips are daisy-chained and in total 256 channels
are read out. We confirmed that the analog output behaves similar to
described in Fig.~\ref{fig:shiftout} (a) where two VA chips were mounted
on the evaluation board that we tested. We also injected a test pulse to
the second VA chip and successfully read out signals from it.

\section{DESIGN OF HYBRID BOARD III and BEAM TEST}\label{hybrid3}

\begin{figure}[t!]
\includegraphics[width=14cm]{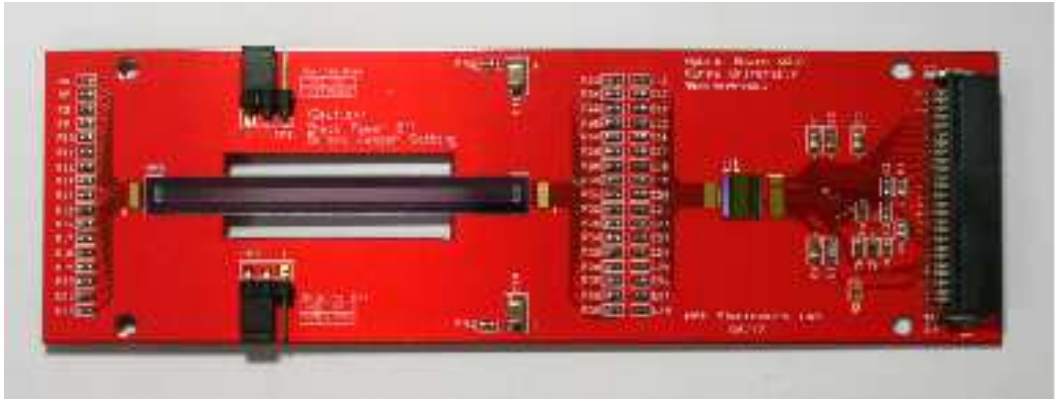}
\caption{
A picture of the hybrid board version 2.0. A 17 channel single-sided 
silicon sensor
is mounted on the left side. Surface-mounted resistor and capacitor arrays are placed
in order to compensate dc current in order to replace the $R/C$ chip.
} 
\label{fig:v20}
\end{figure}

 Based on the experience gained from studies described in 
previous sections, we designed
our hybrid board version 2.0, shown in Fig.~\ref{fig:v20}. This time,
a 17 channel 
single-sided silicon detector (SSD)~\cite{silicon}
is mounted. Also, in order to avoid the
complexity in biasing the detector due to the $R/C$ chip, we decide
to make an array of surface-mount resistors and capacitors to compensate
leakage current. The version 2.0 has in total 6 layers in the PCB
including a power
and a ground plane. One VA chip is hand-mounted with the conductive
epoxy in the
laboratory and all necessary wire-bonding processes are done from the 
company~\cite{lpelectronics}.
Note that there are only 17 channels that are wire-bonded from the sensor
to the VA chip. All electrical tests are carried out and show no trouble.

\begin{figure}[t!]
\includegraphics[width=14cm]{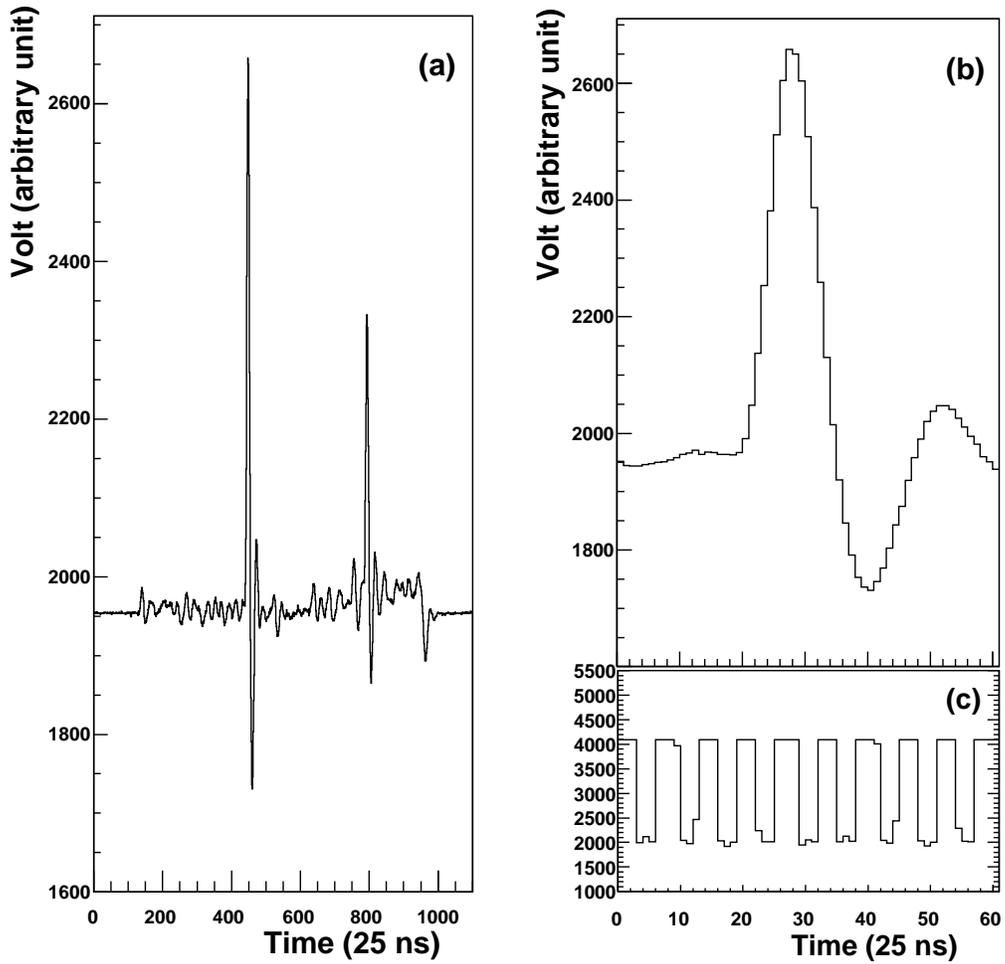}
\caption{
Analog output and clock signals  
from the hybrid board. The analog output from the VA chip
over 128 channels is shown in (a) where the readout clock starts at 100 and ends
at 950 counts in the horizontal axis. A zoomed view of the first peak in (a) is shown in
(b) where an undershooting is clearly visible. The time synchronized clock
signal fed into the VA chip is also shown in (c).   
} 
\label{fig:event}
\end{figure}

 In order to measure real signal from the charged particle, a beam test
is carried out at Korea Institute of Radiological and Medical 
Science (KIRAMS). A small
proton cyclotron with an energy of 35 MeV is used for the test with the
beam current ranging from 0.3 nA to 10 nA. 
Here, we discuss the performance of the electronics only and detailed
performance of the sensor will be addressed in a separate paper.
First, the width of the pedestal we measured at the laboratory
is similar to the level 
at KIRAMS when the proton beam is present,
indicating the electronics does not become noisy
in environment such as the beam area. Then the detector is fully biased and
the signal is read out using the data acquisition system we prepared. 
A trigger signal comes from a 30 ml liquid scintillator that is made of
10 \% of BC501A and 90 \% of mineral oil loading~\cite{hjkim}.
A 12 bit 40 MHz Versamodule Eurocard (VME) flash analog-to-digital converter
(FADC) is used to digitize the analog output signal
from the hybrid board. A VME CPU running a linux operating system
collects data stored
in a 4K word long buffer inside of the FADC. The ROOT~\cite{root} package is  interfaced
with a VME device driver that controls VME slave boards and the data
are stored in a ROOT format for offline analysis. 
A clocked raw output from the hybrid board is shown in Fig.~\ref{fig:event} (a). 
It has two sharp peaks and their positions correspond to the channels that
were wire-bonded to the VA chip. 
These two peaks may correspond to the proton beams detected by the 
silicon strip sensor.
However, we observe undershooting of
the clocked analog outputs. The zoomed view of the peak is in 
Fig.~\ref{fig:event} (b) with the clock signal shown in 
Fig.~\ref{fig:event} (c).
A clear undershooting exists over a clock period. It appears that the
source of the undershooting may be from the driver chip on the PLC board
that converts differential analog output from the VA chip to single-ended
signal. It turns out that 
the speed of the driver chip was much slower than the clock 
speed of the VA chip operation which was 4 MHz for the beam test. 
This may explain a larger noise observed in the signal-to-ratio measurement
shown in the previous section.

\section{CONCLUSIONS}
 A series of electronics boards are fabricated in order to serially read out
small analog signals from high density sensors such as silicon strip 
detectors. A commercially available amplifier is mounted on a homemade 
hybrid board in order to receive analog signals from the detectors. 
Fabrication of the
board, micro-patterning of pads for ultrasonic wire-bonding, and
necessary wire-bonding on the 
hybrid board are carried out and tests show a good performance
of the fabricated board. Also, a compact electronics 
board that provides necessary
current, voltage source, and control logic 
is fabricated in order to communicate with
the hybrid board. 
The linearity of the VA chip is studied in the calibration mode and a good
response is achieved up to 7 MIPs. A VA-test hybrid board is fabricated
and the signal-to-ratio is measured with somewhat worse ENC value then
the specification value.
We expose the electronics and silicon sensors in the
proton beams and verified a good performance of the electronics in the
beam environment. However, the noise value is higher than the nominal
value and there is undershooting in the analog output signal. 
It turns out that the speed of the driver chip on the PLC board may be 
too slow for our application.
These
problems will be examined in future and will be addressed in a separate
paper. The electronics boards discussed in the paper may be used in 
applications including medical imaging and charged particle tracking
system with no thermal dissipation is required. Fabrication of the
hybrid board with a ceramic PCB will be studied in future.

\begin{acknowledgments}
This work is supported by grant No.
R01-2005-000-10089-0
from the Basic Research Program of the Korea Science 
\& Engineering Foundation and supported by the Korea Research Foundation
Grant funded by the Korean Government (MOEHRD) (KRF-2006-C00258
and KRF-2005-070-C00032).
EW is partially supported by a special startup grant from SK Corporation
and by a Korea University Grant.
\end{acknowledgments}

\newpage

\newpage
%%\begin{figure}[htbp]

\newpage
%%\begin{figure}[htbp]

\newpage

\newpage

\newpage

\newpage

\newpage

\newpage

\newpage

\end{document}